\documentclass[conference]{IEEEtran}
\usepackage{graphicx, xcolor, svg, multirow, soul,amssymb, amsmath, amscd, graphicx, epstopdf, hyperref, color, setspace, tensor, url, verbatim,breakurl, caption, subcaption}
\usepackage{caption}
\usepackage{subcaption}

\usepackage{multirow}
\usepackage{amsmath}

\usepackage{listings}
\usepackage{xcolor}
\definecolor{codegreen}{rgb}{0,0.6,0}
\definecolor{codegray}{rgb}{0.5,0.5,0.5}
\definecolor{codepurple}{rgb}{0.58,0,0.82}
\definecolor{backcolour}{rgb}{0.95,0.95,0.92}
\lstdefinestyle{mystyle}{
    backgroundcolor=\color{backcolour},   
    commentstyle=\color{codegreen},
    keywordstyle=\color{magenta},
    numberstyle=\tiny\color{codegray},
    stringstyle=\color{codepurple},
    basicstyle=\ttfamily\footnotesize,
    breakatwhitespace=false,         
    breaklines=true,                 
    captionpos=b,                    
    keepspaces=true,                 
    numbers=left,                    
    numbersep=5pt,                  
    showspaces=false,                
    showstringspaces=false,
    showtabs=false,                  
    tabsize=2
}
\usepackage[export]{adjustbox}

\hypersetup{colorlinks, citecolor=black, filecolor=black, linkcolor=black, urlcolor=black, linktocpage}

\begin{document}
%
\title{Cross Device Federated Intrusion Detector for Early Stage Botnet Propagation in IoT}
%
%

\author{
    \IEEEauthorblockN{Angela Grace Famera\IEEEauthorrefmark{1}, 
    Raj Mani Shukla\IEEEauthorrefmark{2},
    Suman Bhunia\IEEEauthorrefmark{1}
    }
    \IEEEauthorblockA{\IEEEauthorrefmark{1}
    Department of Computer Science and Software Engineering, Miami University, Oxford, Ohio}
    \IEEEauthorblockA{\IEEEauthorrefmark{2}
    Computing and Information Science, Anglia Ruskin University, Cambridge
    }
    \IEEEauthorblockA{
    Email:  fameraag@miamioh.edu, raj.shukla@aru.ac.uk, bhunias@miamioh.edu}
}


\maketitle              
\begin{abstract}
A botnet is an army of zombified computers infected with malware and controlled by malicious actors to carry out tasks such as Distributed Denial of Service (DDoS) attacks. Billions of Internet of Things (IoT) devices are primarily targeted to be infected as bots since they are configured with weak credentials or contain common vulnerabilities. Detecting botnet propagation by monitoring the network traffic is difficult as they easily blend in with regular network traffic. The traditional machine learning (ML) based Intrusion Detection System (IDS) requires the raw data to be captured and sent to the ML processor to detect intrusion. In this research, we examine the viability of a cross-device federated intrusion detection mechanism where each device runs the ML model on its data and updates the model weights to the central coordinator. This mechanism ensures the client's data is not shared with any third party, terminating privacy leakage. The model examines each data packet separately and predicts anomalies. We evaluate our proposed mechanism on a real botnet propagation dataset called MedBIoT. Overall, the proposed method produces an average accuracy of 71\%, precision 78\%, recall 71\%, and f1-score 68\%. In addition, we also examined whether any device taking part in federated learning can employ a poisoning attack on the overall system. 
\end{abstract}

 \section{Introduction}
Computer security plays an imperative role in our technologically advanced society. 65\% of business leaders in 2019 felt their cybersecurity risks were increasing, and the average cost of a malware attack on a company is currently \$2.6 million \cite{varonis}. With the advancement of smart technology, attacks on IoT devices also tripled in the first half of 2019, and the Mirai Distributed Denial of Service (DDoS) worm (the Mirai botnet) was the third most common IoT threat in 2018 \cite{varonis}. By 2025, there will be 41.6 billion connected IoT devices generating 79 zettabytes of data \cite{sectigo}, and damage related to cybercrime is projected to hit \$10.5 trillion annually \cite{varonis}.

Effective techniques to actively detect malicious network traffic are firewalls, Intrusion Detection Systems (IDS), and Intrusion Prevention Systems (IPS). These defense techniques require analyzing raw network traffic data, which can cause privacy concerns when multiple devices and data silos are owned by independent individuals and organizations. As effective as they are, firewalls, IDS, and IPS are not built to conserve data privacy or learn independently from the data they are fed. Federated learning can address both privacy and self-learning. This research aims to create an IDS to detect IoT botnets on a packet-by-packet basis using federated learning before an attack can take place. 

A typical IDS sees all the network traffic coming into a network.
Federated learning is most popularly explained in the medical setting, given its ability to keep patient data confidential between different organizations. But why would we want to keep network data private to the device? Packets hold an array of information that could be used to cause harm if it falls into the wrong hands. We want to use federated learning to aggregate more information from multiple networks.  Our approach allows us to connect to multiple networks and process more data without actually seeing the data. More data from different networks allows us to build better models because they will have more content to learn from. Moreover, the nature of anomaly detection will be more robust. Rather than building a model to recognize ``normal'' traffic and flag everything that does not fit that standard, we also want our model to recognize hard-to-detect malicious traffic. For example, during botnet propagation, it is very easy for that traffic to blend in with normal traffic. While it is good to be able to predict attack traffic, ideally, we would like to take a more proactive approach to recognize anomalies before harm can be done. Especially since the main attack mechanism of botnets is DDoS attacks, there is not much someone can do to mitigate that circumstance in the heat of the moment. 

The main contributions of this research are the following:
\begin{itemize}
    \item We propose a new intrusion detection system mechanism based on federated learning to preserve data privacy. Each device takes part in federated learning by training the model locally, so no data is shared. 
    \item Usage of raw packet data from real and emulated network traffic captures during propagation and C\&C communication for three popular IoT botnets.
    \item We propose an online model analyzing network data on a per-packet basis
    \item We perform the novel feature selection on the network data for efficient anomaly detection \footnote{All source code posted at \url{https://github.com/sbhunia/ml-malware}}. 
    \item We designed a neural network model to identify botnet traffic at its early stages (i.e. pre-attack)
    \item We examined whether poisoning attacks have an impact on model performance. A poisoning attack occurs when an adversary injects bad data into a model. We simulate this kind of attack using label-flipping based on known malware trends.
    \item Overall, the proposed method produces an average accuracy of 71\%, precision 78\%, recall 71\%, and f1-score 68\%.
\end{itemize}

\section{Background \& Related Work}

\subsection{Botnet Overview}
While botnets often serve the hacker today, they were originally developed to assist with the administration of Internet Relay Chat (IRC) servers \cite{Hyslip2015_SurveyOnDetectionTechniques}. IRC is a text-based protocol developed in 1988 used by connected computers for real-time text messaging \cite{radware}. The IRC consisted of five main components: servers, clients, operators, channels, and channel operators \cite{Oikarinen1993_IRC}. 
Botnets developed into an attack mechanism used by cybercriminals to perform various malicious actions, most commonly being Distributed Denial of Service (DDoS) attacks, spam distribution, and network scanning, exploration, and exploitation \cite{Chang2015}. A botnet is a collection of bots connected to and controlled by a C\&C channel \cite{Sergio2013_BotnetsASurvey}. Bots are constituted of host machines, devices, and computers infected by malicious code that enslaves them to the C\&C \cite{Sergio2013_BotnetsASurvey}. The C\&C updates and guides bots to perform the desired task by acting as the communication link between the bots and an individual known as a botmaster \cite{Sergio2013_BotnetsASurvey}. The botmaster's primary purpose is to control the botnet by issuing commands through the C\&C to perform malicious and illegal activities.

The most critical part of a botnet is the C\&C architecture \cite{Sergio2013_BotnetsASurvey}. The C\&C is the only way to control the bots within a botnet and is responsible for their smooth and collective operation. Therefore, if the C\&C was destroyed, the botnet would no longer be able to carry out its intended purpose. The three most common botnet C\&C structures are centralized, decentralized, and hybridized control \cite{Sergio2013_BotnetsASurvey, Gaonkar2020_ASurveyOnBotnetDetectionTechniques}.
Centralized C\&C primarily uses HTTP and IRC based protocols.  Unlike centralized C\&C, decentralized C\&C uses Peer-to-Peer (P2P) protocols in which all the bots are connected. These protocols focus on hiding the C\&C channels, and botmasters can call different bots for different issues. Hybridized botnets use a combination of the centralized and decentralized C\&C structure, often using encryption to hide botnet traffic. 

Botnets have a unique life cycle \cite{Sergio2013_BotnetsASurvey}:
\begin{itemize}
    \item \textbf{Initial Injection:} The host device is infected and becomes a potential bot. During this phase, the attacker may use an array of different infection mechanisms, such as infected files and removal disks or forced downloads of malware from various websites.
    \item \textbf{Secondary Injection:} The infected host runs a program that transforms the device into a bot. Scripts are executed by the infected host that fetches the device's binary code via FTP, HTTP, or P2P protocol. This binary contains the addresses of the machines and may be encoded directly as hard-coded IP addresses or domain names. During this phase, the host becomes a bot.
    \item \textbf{Connection:} Through a process known as rallying, the bot establishes a connection with the C\&C server and becomes part of the botmaster's botnet. This phase occurs every time the host is restarted to let the botmaster know that the device is still able to receive and act on malicious commands. 
    \item \textbf{Performance:} The bot is able to receive commands and perform attacks. The C\&C  enables the botmaster to monitor and control the botnet however seen fit.
    \item \textbf{Maintenance:} The bot's malware is updated for the botmaster to maintain the botnet. Here binary's are updated to ensure a connection with the C\&C remains established. 
\end{itemize}
Methods of propagation include but are not limited to email attachments, infected websites, and previously installed backdoors \cite{Sergio2013_BotnetsASurvey}.

\subsection{Federated Learning }
\label{flprocess}
\begin{figure}%
    \centering
    \begin{subfigure}{0.45\linewidth}
    \includegraphics[width=\linewidth]{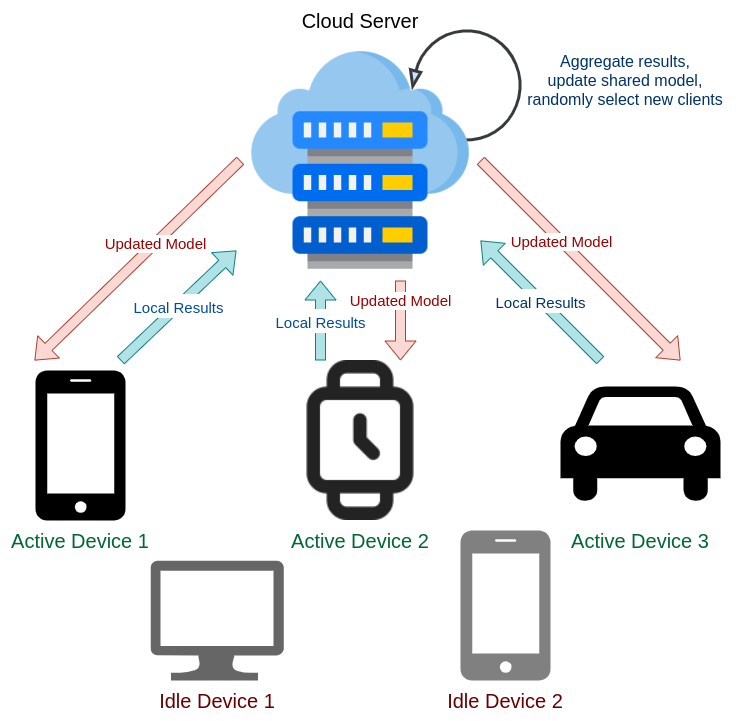}
    \caption{Cros-Device}
    \end{subfigure}
    \begin{subfigure}{0.45\linewidth}
    \includegraphics[width=\linewidth]{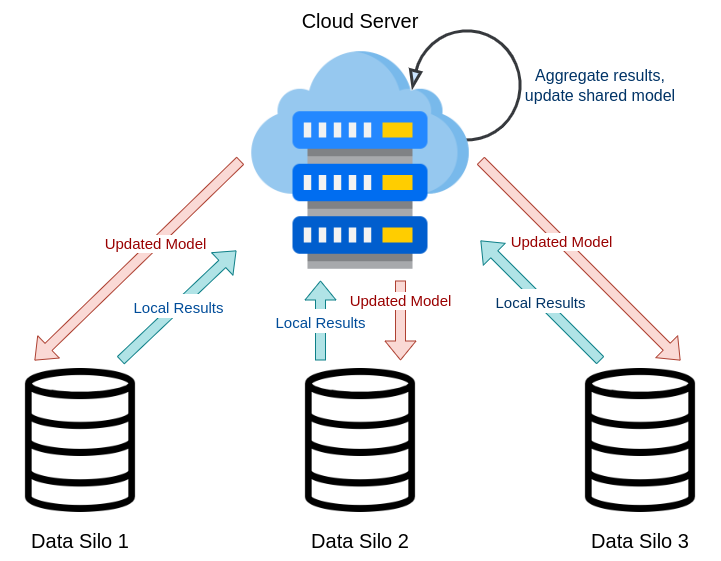} 
    \caption{Cross-silo}
    \end{subfigure}
    \caption{Federated Learning Architectures}%
    \label{fig:fedlearnarch}%
    \vspace{-15pt}
\end{figure}

Federated learning is a machine learning method where multiple clients collectively train a model using their own data in a decentralized manner under the guidance of a central server \cite{rajesh2023take}. In 2016, the term ``Federated Learning'' was coined by McMahan et al. in \cite{McMahan2016_FL} as a learning task is solved by a loose federation of participating devices. Federated learning differs from distributed learning in the sense that a traditional distributed system encompasses distributed computation and storage \cite{Li2020_ReviewOfAppOfFL}. The primary advantage of the federated learning approach is to remove a model's need for direct access to raw training data, resulting in enhanced privacy \cite{McMahan2016_FL}. Its advantages are often described for the medical setting, where medical organizations can train models on patient data without violating privacy laws or exposing clients' medical history. The emphasis on privacy makes the federated learning framework attractive for cybersecurity, given it will inherently protect data security and confidentiality \cite{Ghimire2022}. There are two notable types of federated learning, 1) Cross-device and 2) Cross-silo (see Figure \ref{fig:fedlearnarch}). The following section will discuss each one more in-depth.

In {\em Crossed Device Federated Learning}, The clients are a very large number (i.e. thousands or millions) of mobile, edge, or IoT devices \cite{zhang2021survey}. Here each client stores its own data and is unable to see the data produced by another client, and the data is not independently or identically distributed \cite{zhang2021survey}. There is a central server that supervises training but does not have access to the raw data, only the model parameters returned by the clients. We use the Cross-device approach in our research. We use a labeled dataset and feed it through distributed clients, sharing a simple neural network model for training. We aim to reduce the complexity of our network given the low processing or computing power IoT devices may exhibit. In our simulation, each client acts as an IoT device and stores network packet data. 

{\em Cross-Silo Federated Learning} used data silos which are repositories of data stored in a standalone system controlled by a single department or business unit within an organization . As the name suggests, Cross-silo federated learning trains a model on siloed data, making the clients different organizations or geographically distributed data centers (typically 2 - 100 clients) \cite{rajesh2023take}. 

The steps in the federated learning process are as follows \cite{zhang2021survey}:
\begin{enumerate}
    \item Identifying the problem to be solved with federated learning.
    \item Training data is distributed among clients (simulation environment).
    \item Federated model training begins.
    \begin{enumerate}
        \item The server samples a set of clients.
        \item Selected clients download current model weights and a training program.
        \item Clients locally compute and update the model.
        \item The server collects an aggregate of the device updates.
        \item The server locally updates its model based on the aggregated data computed from the clients.
    \end{enumerate}
    \item The federated model is evaluated after sufficient training.
    \item The model is deployed within the data center or network using a staged rollout.
\end{enumerate}

\subsection{Related Work}
Malware detection and machine learning are widely researched topics. Ghimire et al. compile a comprehensive survey on federated learning for cybersecurity in relation to IoT devices \cite{Ghimire2022}. They present a detailed study on federated learning models used for cybersecurity and their associated performance metrics and challenges. They discuss the limitations of federated learning in the IoT space, such as limited device memory, battery power, and computing power. In summary, it appears to be the most recent survey of all things federated learning in relation to cybersecurity and IoT devices. 

Galvez et al. present a malware classifier leveraging federated learning for Android applications called LiM (`Less is More') \cite{Galvez2020}. LiM uses a safe semi-supervised learning ensemble (SSL) to maximize accuracy with respect to a baseline classifier on the cloud by ensuring unlabeled data does not worsen the performance of a fully supervised classifier. Prior to \cite{Galvez2020}, Hsu et al. present a privacy-preserving federated learning (PPFL) android malware detection system \cite{Hsu2020}. Implementing the PPFL using support vector machines, the researchers demonstrate its feasibility using an Android malware dataset by the National Institute of Information and Communication Technology (NCIT) containing over 87,000 APK files from the Opera Mobile Store. Zhao et al. propose a multi-task deep neural network in federated learning (MT-DNN-FL) to perform network anomaly detection tasks, VPN (Tor) traffic recognition tasks, and traffic classification tasks simultaneously \cite{Zhao2019}. 

Rey et al. use federated learning to detect malware in IoT devices \cite{Rey2022}. The researchers perform both supervised and unsupervised federated learning using N-BaIoT, a dataset compiled in 2018 containing attack data from nine IoT devices infected by Mirai and Bashlite. They achieve accuracies, True Positive Rates (TPR), and True Negative Rates (TNR) of +99\% in their Multi-Epoch and Mini-Batch supervised models. There are a few key differences between Rey et al. and our research. Two of the most important are: first, we are using an entirely different dataset; and, second,  we are examining our instances on a packet-by-packet basis, similar to how an intrusion detection system would. We do not use statistics collected from the packet streams but rather examine each packet and its contents independently. The analysis of per-packet makes the proposed IDs much faster as opposed to using the stream information. 


\section{Proposed Cross-Device FL Intrusion Detection (ID) Architecture}
\label{proposedarch}

\begin{figure}
    \begin{subfigure}{0.48\linewidth}
        \centering    
        \includegraphics[width=\linewidth]{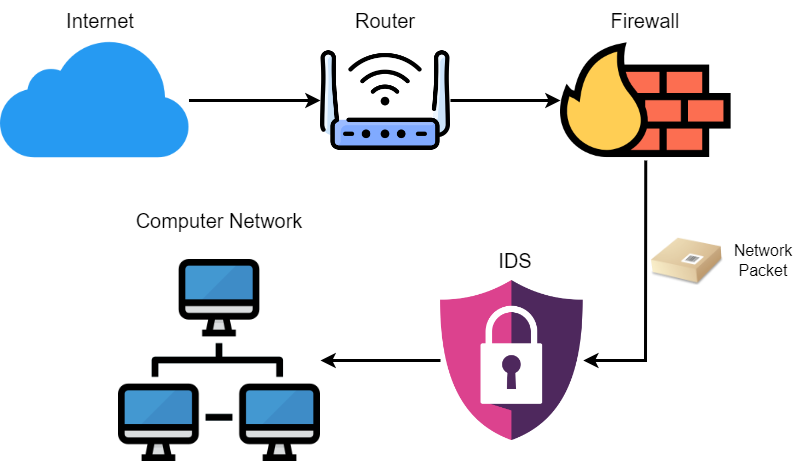}
        \caption{Standard IDS Setup}
        \label{fig:standardIDS}
    \end{subfigure}
    \begin{subfigure}{0.48\linewidth}
        \centering  
        \includegraphics[width=\linewidth]{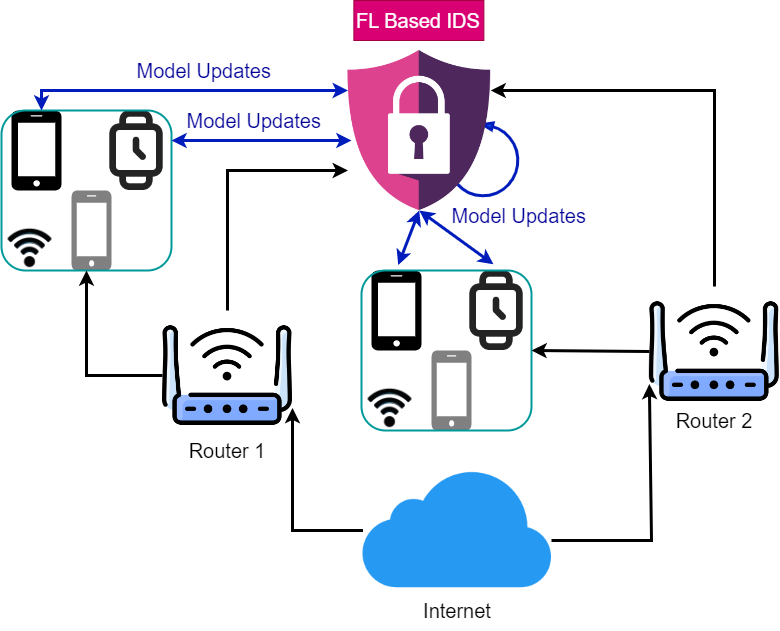}
        \caption{Federated IDS Setup}
        \label{fig:federatedIDS}
    \end{subfigure}
    \caption{IDS architecture}
    \vspace{-15pt}
\end{figure}

Figure \ref{fig:standardIDS} displays a general IDS as it is most widely used in the market. When a router connects to the internet, a firewall is commonly put up to block specified malicious traffic. An IDS can be placed before or after a firewall, but it is optimal to put it after so the firewall takes the brunt of the action, and the IDS can catch malicious stragglers the firewall may miss. Then, whatever traffic passes through the IDS is allowed to enter the private network. 

Anomaly-based IDS solutions model ``normal'' behavior within the system, labeling packets as potential threats if they present anomalous behavior. This is beneficial for a closed or private organizational network where there could be such a thing as ``normal'' traffic behavior, but this becomes difficult to pinpoint as open networks and IoT devices grow in size. We propose a slightly different architecture.

\subsection{Detecting Botnet Propagation}
\label{detectingbotnetprop}
The phases of the botnet lifecycle are initial injection, secondary injection, connection, performance, and maintenance. During the initial and secondary injection, the host device is infected by either weak credentials or system vulnerabilities. For example, Mirai would infect devices by performing a brute force attack at guessing common or insecure credentials, such as ``admin'' and ``admin'' for username and password. Mirai had a list of 60 common usernames and passwords, and would randomly to chose 10 from that list. Bashlite would infect devices via Shellshock, a bash vulnerability that allowed the execution of arbitrary commands to gain unauthorized access when concatenated to the end of function definitions. Torii is more sophisticated in the sense it will download different binary payloads based on the architecture of the targeted device. More sophisticated botnets are sneaky when infecting devices, and it is difficult to tell when a device has been infected. After infection, the device is then connected to the C\&C, and it awaits instruction. Once it receives the instruction to do so, it can then carry out attacks. 

The ability to detect the early botnet phases before receiving instruction from the C\&C to attack is important. We believe analyzing the individual network packets may be one of the best ways to do this. If we can identify communication traffic, we can more likely identify the C\&C and put up defenses against it. To analyze the individual packets, we examine the information stored in the packet headers.  We believe a system that analyzes on a per-packet basis is a beneficial way to detect discrepancies between normal and malicious traffic.

\subsection{Federated Machine Learning Approach}
\label{holisticfedml}
Figure \ref{fig:federatedIDS} displays a holistic overview of our proposed architecture. The components of the architecture are listed below:
\begin{itemize}
    \item \textbf{Federated IDS:} The intrusion detection systems that utilizes federated learning to differentiate between anomalous and normal network data on a per-packet basis.
    \item \textbf{Devices:} Unlike the standard IDS system, our IDS would be able to connect to multiple devices on multiple networks. The devices would train our model, then report the model weights back to the IDS without having to share data.
    \item \textbf{Internet Routers:} What allows the devices to connect to the internet.
\end{itemize}
We want to use federated learning because it conserves data privacy. In theory, our IDS would be connected to different routers, therefore being connected to different networks. The central IDS or server will distribute its model to the selected devices, and each device will run the model on its own network data. Devices would not have to share their network data with other devices or the IDS. To avoid exposing private information across devices or through the internet, all data remains local to the device, and only the model weights are updated and adjusted. We take the approach of trying to differentiate between ``normal'' and ``anomalous'' traffic since normal traffic may be hard to define in larger networks that are open or have many devices connected to them. We propose training our model on pre-attack data to help it focus on learning the discrepancies between malicious and benign network traffic when they are hard to tell apart. 

\subsection{Poisoning Attack by Participant Node}
Poisoning attacks happen when a hacker tries to inject fake training data into a model to reduce or hinder its performance. There are four main categories of poisoning attacks: 1) logic corruption, 2) data manipulation, 3) data injection, and 4) Domain Name System (DNS) cache poisoning \cite{posion}. Logic corruption is where the attacker changes the logic of the systems to disrupt how the system learns \cite{posion}. Data manipulation is where the attacker manipulates the data rather than the logic \cite{posion}. Data injection is when fake data is inserted into the actual dataset to skew model results and weaken outcomes \cite{posion}. Last, DNS cache poisoning is when the attacker corrupts DNS data, causing the name server to return incorrect results \cite{posion}. The type of poisoning attack we try to replicate is a data manipulation attack. Since we are using supervised learning, we emulate an attack as if a malicious user switched the malware labels to benign to cover up bad traffic. We tried to be sophisticated with our approach using label-flipping. We researched the most popular source port among the malicious data in the dataset and found port 23 to be on top. As a result, we flipped all the labels from 1 to 0 for packets with a source port equal to 23 in the first client.

\subsection{Feature Selection}

We use MedBIoT  data focusing on the early stages of botnet deployment: propagation and C\&C communication \cite{MedBIot}. In this dataset, a combination of real and emulated devices are used, making up for a total of 83 devices. The real devices include a Sonoff Tasmota smart switch, TPLink smart switch, and a TPLink light bulb. The Emulated devices include locks, switches, fans, and lights. These devices are infected with the Mirai, Bashlite, and Torii botnet malware. The packet attributes we initially analyzed using Wireshark are :
frame.encap\_type, 
frame.time, 
frame.time\_epoch, 
frame.offset\_shift, 
frame.time\_delta, 
frame.time\_delta\_displayed, 
frame.time\_relative, 
frame.number, 
frame.len, 
frame.cap\_len, 
frame.marked, 
frame.ignored, 
frame.protocols, 
frame.coloring\_rule.name, 
eth.dst, 
eth.src, 
eth.type, 
ip.dsfield, 
ip.len, 
ip.id, 
ip.flags, 
ip.ttl, 
ip.proto, 
ip.checksum, 
ip.checksum.status, 
ip.src, 
ip.dst, 
tcp.srcport, 
tcp.dstport, 
tcp.stream, 
tcp.len, 
tcp.seq, 
tcp.seq\_raw, 
tcp.nxtseq, 
tcp.ack, 
tcp.ack\_raw, 
tcp.hdr\_len, 
tcp.flags, 
tcp.window\_size\_value, 
tcp.window\_size, 
tcp.window\_size\_scalefactor, 
tcp.checksum, 
tcp.checksum.status, 
tcp.urgent\_pointer, 
tcp.time\_relative, 
tcp.time\_delta, 
tcp.analysis.bytes\_in\_flight, 
tcp.analysis.push\_bytes\_in\_flight, 
udp.srcport, 
udp.dstport, 
udp.length, 
udp.checksum, 
udp.time\_relative, 
udp.time\_delta.

\begin{table}[t]
\centering
\caption{ Reason for Removing Features}
\label{tab:resonremove}
\scriptsize
\begin{tabular}{ll}
\hline
\textbf{Key} & \textbf{Reason for Removal} \\ \hline
frame.encap\_type & Lack of Variability \\  
frame.time & Date/Time \\  
frame.time\_epoch & Continuous \\  
frame.offset\_shift & Lack of Variability \\  
frame.time\_delta & Continuous \\  
frame.time\_delta\_displayed & Interchangeable (frame.time\_delta) \\  
frame.time\_relative  & Continuous \\  
frame.number & Continuous \\  
frame.cap\_len & Interchangeable (frame.len) \\  
frame.marked & Lack of Variability \\  
frame.ignored & Lack of Variability \\  
frame.coloring\_rule.name  & N/A \\  
eth.dst & Categorical \\  
eth.src & Categorical \\  
eth.type  & Lack of Variability \\  
ip.dsfield & Lack of Variability \\  
ip.id & Identification \\  
ip.checksum & Variability \\  
ip.checksum.status & Lack of Variability \\  
tcp.stream & Identification \\  
tcp.seq & Variability \\  
tcp.seq\_raw & Variability \\  
tcp.nxtseq & Variability \\  
tcp.ack & Variability \\  
tcp.ack\_raw & Variability \\  
tcp.checksum & Variability \\  
tcp.checksum.status & Lack of Variability \\  
tcp.urgent\_pointer & Lack of Variability \\  
udp.srcport & Nulls  \\  
udp.dstport & Nulls  \\  
udp.length & Nulls  \\  
udp.checksum & Nulls  \\  
udp.time\_relative & Nulls  \\  
udp.time\_delta  & Nulls  \\  \hline
\end{tabular}
    \vspace{-15pt}
\end{table}

We first generated a large, singular dataset from the bulk data containing all the features, then randomly selected a little over 3,700 rows to produce a subset dataset to gain a more holistic view of the values each feature contained. After analyzing these features in the subset, we decided to discard a handful of them as they did not have variable information.

Now that we had the features we wanted, we then added a few additional features to use as labels.  The first feature we modified was frame.protocols. These were long character strings, so we split up the string and kept the last element. For example, if the data point was ``eth:ethertype:ip:tcp'' we manipulated it such that only ``tcp" was kept. After splitting the protocol, we then created various labels. The first label we created was used to display whether a packet was malicious or not. After creating is\_malware, we then created a label differentiating the botnet's action. To do this, we first determined whether the file was malicious or legitimate. Then, based on the file name, we labeled it as ``spread'' for propagation traffic, and ``cc'' for C\&C communication traffic. Note that all the Torii files only housed propagation data given the researchers did not want to risk communication with Torii's real C\&C server. Once we created a feature for the botnet phase, we then created a feature for the device type used. The types of devices included a fan, switch, lock, light, and two raspberry pis.

After creating these additional labels, we transformed each fine-grained csv and added ``mod'' in the name. For example, mirai\_mal\_CC\_lock.csv became \\ mirai\_mal\_CC\_lock\_mod.csv. Once the modified datasets were generated with the additional labels, we then combined these files based on device type, but still differentiated between malicious and legitimate traffic. After combining the datasets by device and traffic type, we then selected 2,000 random rows from each dataset and generated a singular dataset containing a total of 24,000 rows. By doing all this, we tried to keep an even amount of data as well as ensure the dataset was an appropriate size to process efficiently. We then proceeded to do some minimal data preprocessing.

We removed all of the features containing null values since it would be difficult to fill this data in.  We ended up with a mildly unbalanced dataset containing 23,793 samples in total out of the original 24,000. Out of the 23,793 samples, 11,942 were labeled as malware, and 11,851 were labeled as benign. While we tried to keep things as balanced as possible, we didn't stress too much about this because we inevitably would have to shuffle the data and distribute it randomly amongst the clients. Each client, therefore, will more likely have unbalanced data when training their local model. In the real world, it is also unlikely each client will be processing balanced network traffic. 

\begin{figure*}
    \centering    
    \begin{subfigure}{0.48\linewidth}
        \includegraphics[width=\linewidth]{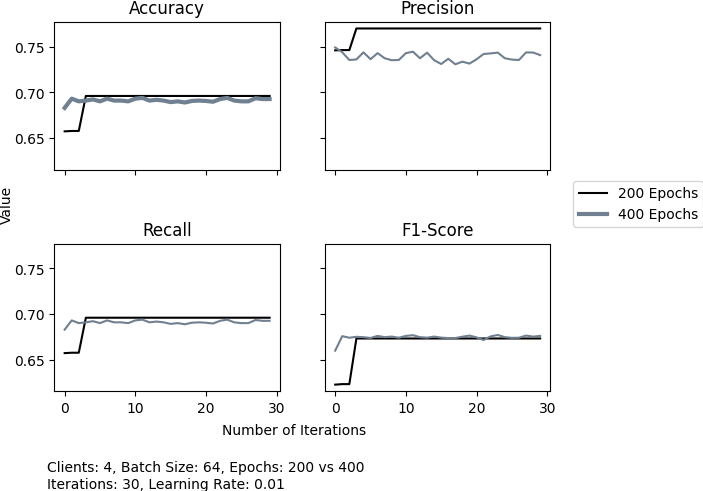}
        \caption{Learning Rate = 0.01}
        \label{fig:train_200_400_0.01}   
    \end{subfigure}
    \begin{subfigure}{0.48\linewidth}
        \includegraphics[width=\linewidth]{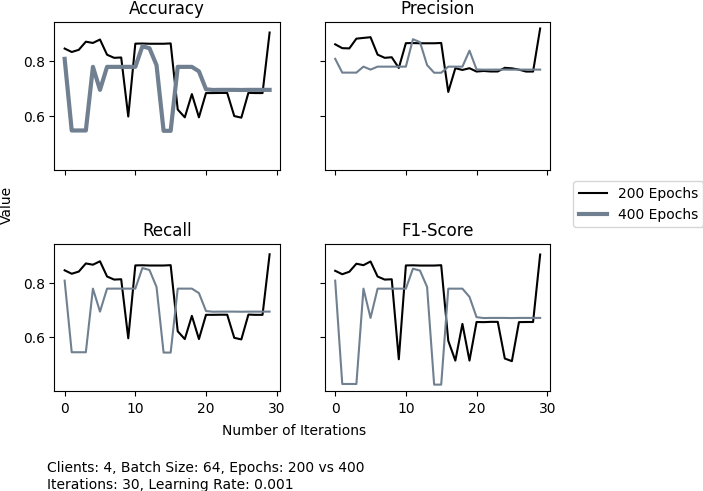}
        \caption{Learning Rate = 0.001}
        \label{fig:train_200_400_0.001}
    \end{subfigure}
    \caption{Scores Produced While Training 200 vs 400 Epochs}
    \vspace{-15pt}
\end{figure*}

\section{Results}
\label{results}
We ran our model using various number if clients, epochs, learning rate, and iterations. The following sections discusses  the results obtained using these hyperparameters and metrics.

\subsection{Hyperparameter: Epochs (200 vs 400)}

Using four clients, a batch size of 64, 30 iterations, and a learning rate of 0.01, we found a negligible difference between 200 and 400 epochs. At the end of the last iteration of training, 200 epochs resulted in an accuracy of 70\%, precision of 77\%, recall of 70\%, and f1 score of 67\%, while 400 epochs resulted in an accuracy of 69\%, precision of 74\%, recall of 69\%, and an f1 score of 68\% (see Figure \ref{fig:train_200_400_0.01}). After running our test data through the model, 200 epochs resulted in an accuracy of 78\% and a loss of 54\%, while 400 epochs resulted in an accuracy of 69\% and a loss of 56\%.

We then tried comparing 200 vs 400 epochs using a smaller learning rate of 0.001. At the end of the last iteration of training, 200 epochs resulted in an accuracy of 91\%, precision of 92\%, recall of 91\%, and f1 score of 90\%, while 400 epochs resulted in an accuracy of 69\%, precision of 77\%, recall of 69\%, and an f1 score of 67\% (see Figure \ref{fig:train_200_400_0.001}). After running our test data through the model, 200 epochs resulted in an accuracy of 67\% and a loss of 57\%, while 400 epochs resulted in an accuracy of 69\% and a loss of 62\%.

While 400 epochs have a slightly higher accuracy during testing for a learning rate of 0.001, it under-performs in almost all other regards compared to 200 epochs. At a learning rate of 0.001, 200 epochs also seem to fluctuate more for 30 iterations, suggesting we may need to increase this parameter. Since there was no significant improvement in increasing the epoch size past 200 for either learning rate, all further testing was performed using 200 epochs to improve time efficiency. 

\subsection{Model Performance After Ten Runs}
\label{benign_performance}

\begin{table}
\centering
\caption{Scores Across 10 Experiments}
\label{tab:scores_10_attempts}
\tiny
\begin{tabular}{|c|cccc|cc|}
\hline & \multicolumn{4}{c|}{Training} & \multicolumn{2}{c|}{Testing} \\ \hline Experiment & \multicolumn{1}{c|}{Accuracy} & \multicolumn{1}{c|}{Precision} & \multicolumn{1}{c|}{Recall} & F1-Score & \multicolumn{1}{c|}{Accuracy} & Loss \\ 

\hline 1  & \multicolumn{1}{c|}{67\%}  & \multicolumn{1}{c|}{79\%} & \multicolumn{1}{c|}{67\%} & 63\%  & \multicolumn{1}{c|}{67\%} & 58\% \\ 

\hline 2  & \multicolumn{1}{c|}{66\%} & \multicolumn{1}{c|}{79\%} & \multicolumn{1}{c|}{66\%} & 62\% & \multicolumn{1}{c|}{66\%} & 63\% \\ 

\hline 3 & \multicolumn{1}{c|}{82\%} & \multicolumn{1}{c|}{82\%} & \multicolumn{1}{c|}{82\%} & 82\% & \multicolumn{1}{c|}{70\%} & 61\% \\ 

\hline 4 & \multicolumn{1}{c|}{65\%} & \multicolumn{1}{c|}{74\%} & \multicolumn{1}{c|}{65\%} & 62\% & \multicolumn{1}{c|}{64\%} & 60\% \\ 

\hline 5 & \multicolumn{1}{c|}{78\%} & \multicolumn{1}{c|}{78\%} & \multicolumn{1}{c|}{78\%} & 78\% & \multicolumn{1}{c|}{78\%} & 49\% \\ 

\hline 6 & \multicolumn{1}{c|}{70\%} & \multicolumn{1}{c|}{71\%} & \multicolumn{1}{c|}{70\%} & 69\% & \multicolumn{1}{c|}{70\%} & 57\% \\ 

\hline 7 & \multicolumn{1}{c|}{54\%} & \multicolumn{1}{c|}{76\%} & \multicolumn{1}{c|}{54\%} & 43\% & \multicolumn{1}{c|}{64\%} & 63\% \\

\hline 8 & \multicolumn{1}{c|}{74\%} & \multicolumn{1}{c|}{79\%} & \multicolumn{1}{c|}{74\%} & 72\% & \multicolumn{1}{c|}{73\%} & 53\% \\ 

\hline 9 & \multicolumn{1}{c|}{54\%} & \multicolumn{1}{c|}{76\%} & \multicolumn{1}{c|}{54\%} & 43\% & \multicolumn{1}{c|}{54\%} & 66\% \\ 

\hline 10 & \multicolumn{1}{c|}{81\%} & \multicolumn{1}{c|}{81\%} & \multicolumn{1}{c|}{81\%} & 81\% & \multicolumn{1}{c|}{81\%} & 51\% \\ 

\hline \textbf{Average} & \multicolumn{1}{c|}{\textbf{69\%}} & \multicolumn{1}{c|}{\textbf{78\%}} & \multicolumn{1}{c|}{\textbf{69\%}} & \textbf{66\%} & \multicolumn{1}{c|}{\textbf{69\%}}  & \textbf{58\%} \\ \hline
\end{tabular}
    \vspace{-15pt}
\end{table}

\subsubsection{Training}
\label{training_benign}
Using four clients, a batch size of 64, 50 iterations, and a learning rate of 0.001, we ran our model ten times to account for some randomness and gain an understanding of the model's overall performance. The scores produced by each experiment are shown in Table \ref{tab:scores_10_attempts}.

Based on the  Table \ref{tab:scores_10_attempts}, E3 and E10 produce accuracies over 80\%, while E7 and E9 produce accuracies below 60\%. E1, E2, and E4 produce accuracies between 60\%-70\%, and E5, E6, and E8 produce accuracies between 70\%-80\%. When looking at the graph, E4, E7, and E9 all exhibit a sharp decrease in accuracy prior to the 20th iteration and stay stagnant on out. E1, E6, E8, and E10 stay relatively consistent throughout the iterations, and E2, E3, and E5 show an increase by the end of the iterations. 

Based on the Table \ref{tab:scores_10_attempts}, E3 and E10 produce precision values over 80\%, while all other experiments produce precision values between 70\%-80\%. When looking at the graph, E4, E7, and E9 all exhibit a sharp decrease in precision prior to the 20th iteration and remain stagnant on out. E6, E8, and E10 stay relatively consistent throughout the iterations, and E1, E2, and E5 display a moderate increase in precision. E3 fluctuates, then stops close to where it started during the first iteration. 

\begin{figure}
    \centering    
    \includegraphics[width=\linewidth]{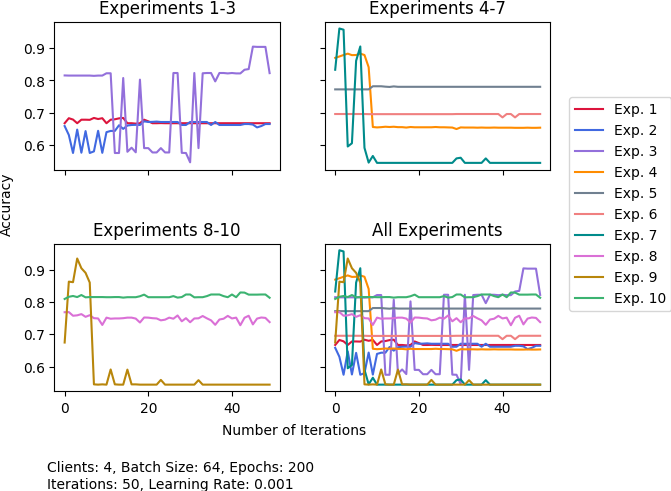}
    \caption{Accuracy Produced Through 10 Repetitions of Training}
    \label{fig:acc_train_10_cases}
    \includegraphics[width=0.6\linewidth]{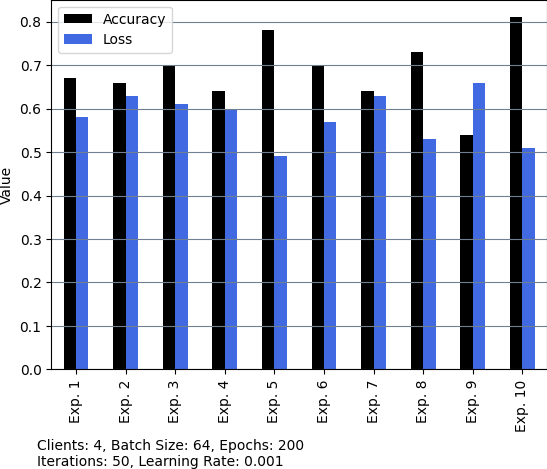}
    \caption{Accuracy and Loss Through 10 Repetitions of Testing}
    \label{fig:test_10_cases}
    \vspace{-15pt}
\end{figure}

The graphs for the recall and f1-scores show that these values produce the same shape as the accuracy for each experiment in Figure \ref{fig:acc_train_10_cases}, with recall producing the same scores as accuracy by the end of the 50th iteration, and the f1 score being slightly less than accuracy. 

Overall, throughout the ten iterations, our model produces an average accuracy of 69\%, precision of 78\%, recall of 69\%, and f1 score of 66\%. We see less fluctuation in precision than accuracy, and our precision values are higher than our accuracy values. Our recall is the same as our accuracy, and our f1 score is less than accuracy, precision, and recall. This fluctuation is likely due to randomness. 

\subsubsection{Testing}
\label{testing_beneign}
After running the testing set through our model, Figure \ref{fig:test_10_cases} shows that 60\% of the experiments fall between 60\%-70\% accuracy (E1, E2, E3, E4, E6, and E7). 20\% of the experiments fall between 70\%-80\% accuracy (E5 and E8), and only 10\% of the experiments achieved over 80\% accuracy (E10). The remaining 10\% (E9) fell between 50\%-60\%. 

Figure \ref{fig:test_10_cases} also shows that while we have relatively high loss values, 90\% of the experiments (with the exception of E9) have a loss value less than that of its associated accuracy. Out of the 60\% of experiments that fall between 60\%-70\% accuracy, half of them have a loss value within the same range, and half have a loss value less than 60\%. We see the greatest disparity between accuracy and loss E5 and E10 with a difference of 29 and 30 respectively. 

On average, we achieve an accuracy of 69\% and a loss of 58\%. Meaning on average, we achieve the same accuracy during testing as we did during training (see Table \ref{tab:scores_10_attempts}). 


\subsection{Model Performance After Ten Runs (Poisoning Attack)}

\begin{table}
\centering
\caption{Scores Across 10 Experiments (Poisoning Attack)}
\label{tab:scores_10_attempts_label_flip}
\tiny
\begin{tabular}{|c|cccc|cc|}
\hline & \multicolumn{4}{c|}{Training} & \multicolumn{2}{c|}{Testing}         \\ \hline 
Experiment  & \multicolumn{1}{c|}{Accuracy} & \multicolumn{1}{c|}{Precision} & \multicolumn{1}{c|}{Recall} & F1-Score & \multicolumn{1}{c|}{Accuracy} & Loss \\ \hline

1 & \multicolumn{1}{c|}{73\%} & \multicolumn{1}{c|}{74\%}  & \multicolumn{1}{c|}{73\%}   & 72\% & \multicolumn{1}{c|}{89\%} & 42\% \\ \hline

2 & \multicolumn{1}{c|}{69\%} & \multicolumn{1}{c|}{77\%}  & \multicolumn{1}{c|}{69\%}   & 67\% & \multicolumn{1}{c|}{69\%} & 62\% \\ \hline

3 & \multicolumn{1}{c|}{53\%} & \multicolumn{1}{c|}{76\%}  & \multicolumn{1}{c|}{53\%}   & 40\% & \multicolumn{1}{c|}{53\%} & 66\% \\ \hline

4 & \multicolumn{1}{c|}{73\%} & \multicolumn{1}{c|}{74\%}  & \multicolumn{1}{c|}{73\%}   & 72\% & \multicolumn{1}{c|}{73\%} & 52\% \\ \hline

5 & \multicolumn{1}{c|}{69\%} & \multicolumn{1}{c|}{72\%}  & \multicolumn{1}{c|}{69\%}   & 68\% & \multicolumn{1}{c|}{69\%} & 64\% \\ \hline

6 & \multicolumn{1}{c|}{87\%} & \multicolumn{1}{c|}{88\%}  & \multicolumn{1}{c|}{87\%}   & 87\% & \multicolumn{1}{c|}{86\%} & 47\% \\ \hline

7 & \multicolumn{1}{c|}{73\%} & \multicolumn{1}{c|}{74\%}  & \multicolumn{1}{c|}{73\%}   & 72\% & \multicolumn{1}{c|}{73\%} & 58\% \\ \hline

8 & \multicolumn{1}{c|}{92\%} & \multicolumn{1}{c|}{93\%}  & \multicolumn{1}{c|}{92\%}   & 92\% & \multicolumn{1}{c|}{65\%} & 60\% \\ \hline

9 & \multicolumn{1}{c|}{69\%} & \multicolumn{1}{c|}{76\%}  & \multicolumn{1}{c|}{69\%}   & 67\% & \multicolumn{1}{c|}{72\%} & 60\% \\ \hline

10 & \multicolumn{1}{c|}{56\%} & \multicolumn{1}{c|}{76\%}  & \multicolumn{1}{c|}{56\%}   & 45\% & \multicolumn{1}{c|}{54\%} & 67\% \\ \hline

\textbf{Average} & \multicolumn{1}{c|}{\textbf{71\%}} & \multicolumn{1}{c|}{\textbf{78\%}}  & \multicolumn{1}{c|}{\textbf{71\%}}   & \textbf{68\%} & \multicolumn{1}{c|}{\textbf{70\%}} & \textbf{58\%}\\ \hline
\end{tabular}
    \vspace{-15pt}
\end{table}

To label-flip, we took all of the rows in the first client with a source port equal to 23 and changed the label from 1 to 0. Using one corrupted client, three clean clients, a batch size of 64, 50 iterations, and a learning rate of 0.001, we ran our model ten times to account for some randomness and gain an understanding of the model's performance. The scores produced by each experiment are shown in Table \ref{tab:scores_10_attempts_label_flip}.

\subsubsection{Training}
\begin{figure}
    \centering    
    \includegraphics[width=\linewidth]{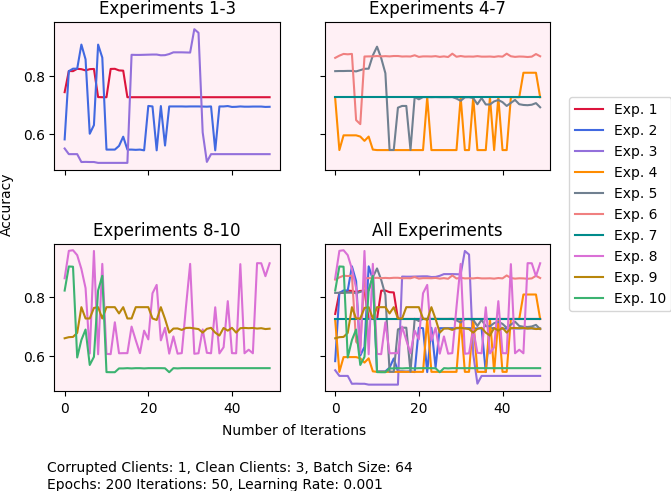}
    \caption{Accuracy Produced Through 10 Repetitions of Training (Poisoning Attack)}
    \label{fig:acc_train_label_flip_three_cases}
    \includegraphics[width=0.7\linewidth]{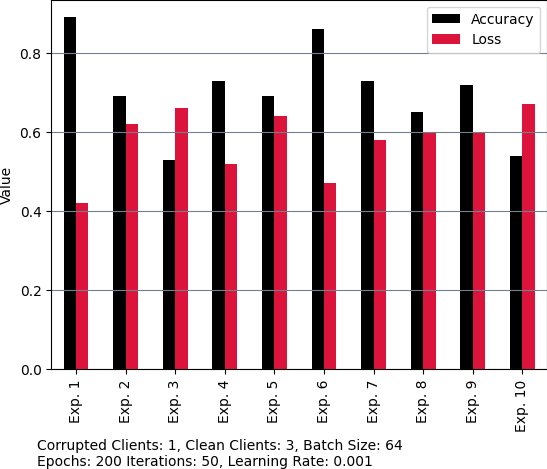}
    \caption{Accuracy and Loss Through 10 Repetitions of Testing (Poisoning Attack)}
    \label{fig:test_label_flip_three_cases}
    \vspace{-15pt}
\end{figure}

Figure \ref{fig:acc_train_label_flip_three_cases} displays the progression of accuracy through 50 iterations for the 10 experiments. Based on the graph and Table \ref{tab:scores_10_attempts_label_flip}, E6 and E8 produce accuracies over 80\% and 90\% respectively, while E3 and E10 produce accuracies below 60\%. E2, E5, and E9 produce accuracies between 60\%-70\%, and E1, E4, and E7 produce accuracies between 70\%-80\%. When looking at the graph, E1, E2, E5, and E10 produce a noticeable decrease in accuracy prior to the 20th iteration. E4, and E8 experience rampant fluctuation throughout the iterations, While E3 and E9 increase and then decrease again in a less chaotic manner. E6 and E7 stay relatively consistent throughout the iterations. Unlike the scores produced in Section \ref{training_benign}, while slightly higher accuracies seemed to be achieved, there is noticeably more fluctuation throughout the learning process for the model. 

Based on Table \ref{tab:scores_10_attempts_label_flip}, E6 and E8 produce precision values over 80\% and 90\% respectively, while all other experiments produce precision values between 70\%-80\%. Looking at the graph, E1, E2, E5, E9, and E10 all experience a decrease in precision throughout the iterations. E3 and E8 experience a minor increase in precision while fluctuating more than the other experiments. E4, E6, and E7 experience little to no fluctuation in precision values throughout the iterations. 

The recall and f1 scores produce the same shape as the accuracy for each experiment in Figure \ref{fig:acc_train_label_flip_three_cases}, with recall producing the same scores as accuracy by the end of the 50th iteration, and the f1 score being slightly less than accuracy (with the exception of E6 and E8, which produce the same as accuracy and recall). 

Overall, throughout the 10 iterations, our model produces an average accuracy of 71\%, precision of 78\%, recall of 71\%, and f1 score of 68\%. Compared to our results in Section \ref{training_benign}, label-flipping achieves higher accuracy, the same precision, higher recall, and a higher f1 score. We see less fluctuation in precision than accuracy, and our precision values are higher than our accuracy values. Our recall is the same as our accuracy, and our f1 score is mostly less than all accuracy, precision, and recall. 

\label{training_mal}


\subsubsection{Testing}
\label{testing_mal}
After running the testing set through our model, Figure \ref{fig:test_label_flip_three_cases} shows 30\% of the experiments (E2, E5, and E8) produce an accuracy between 60\%-70\%. 30\% of the experiments fall between 70\%-80\% accuracy (E4, E7, and E9), and 20\% of the experiments achieved over 80\% accuracy (E1 and E6). The remaining 20\% (E3 and E10) fell between 50\% and 60\%. Compared to testing results in Section \ref{testing_beneign}, label-flipping seems to produce more sporadic results given 20\% as opposed to 10\% of the experiments produce accuracies over 80\%, while 20\%, as opposed to 10\%, produce accuracies below 60\%. 

Figure \ref{fig:test_label_flip_three_cases} also shows that while we have relatively high loss values, 80\% of the experiments (excluding E3 and E10) have a loss value less than that of its associated accuracy. Most of the loss values fall between 60\%-70\% (E2, E3, E5, E8, E9, and E10). The other 40\% of the experiments (E1, E4, E6, and E7) produce loss values below 60\%. 20\% of the experiments (E3 and E10) produce loss values higher than their associated accuracy, which is twice as much as the experiments from Section \ref{testing_beneign}. Last, we see the greatest disparity between accuracy and loss for E1 and E6 with a difference of 47 and 39 respectively. On average, we achieve an accuracy of 70\% and a loss of 58\%. Meaning on average, we achieve a slightly lower accuracy during testing than we did during training (see Table \ref{tab:scores_10_attempts_label_flip}). Compared to Section \ref{testing_beneign}, we have slightly higher accuracy and the same loss.

\begin{figure}%
    \centering
    \subfloat[\centering Model Performance]{{\includegraphics[width=7cm]{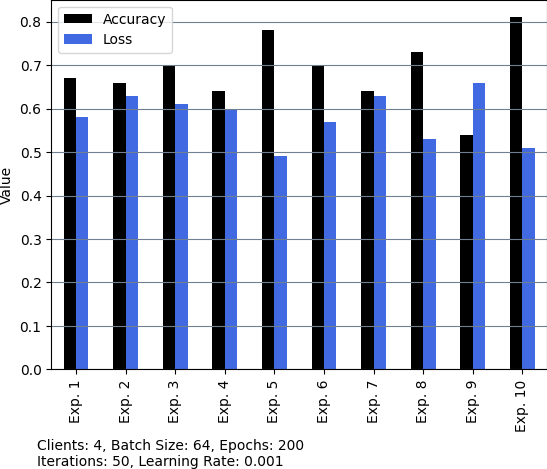} }}%
    \qquad
    \subfloat[\centering Model Performance During Poisoning Attack]{{\includegraphics[width=7cm]{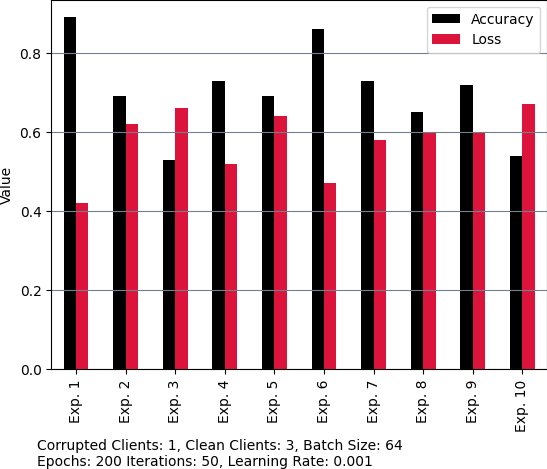} }}%
    \caption{Difference in Model Performances}%
    \label{fig:modelcomparison}%
    \vspace{-15pt}
\end{figure}
%
%
%

\section{Conclusion}
We proposed an intrusion detection system that utilizes federated learning to preserve data privacy amongst devices. We analyze raw packet data from legitimate botnet traffic. Sampling packets from the propagation and C\&C communication phase, we propose an online model to differentiate between malicious and benign traffic on a per-packet basis without allowing the clients to share raw network data. We also examined whether poisoning attacks have an impact on model performance. The proposed method has resulted in an overall accuracy of 71\%, precision 78\%, recall 71\%, and f1-score of 68\%. In future, we plan to implement the proposed federated learning based intrusion detection system in the hardware systems for testing purpose.


\bibliographystyle{ieeetr}
\bibliography{Thesis}

\begin{thebibliography}{10}

\bibitem{varonis}
R.~Sobers, ``134 cybersecurity statistics and trends for 2021: Varonis,'' Mar
  2021.

\bibitem{sectigo}
Sectigo, ``Evolution of iot attacks: An interactive infographicd.''
  \url{https://sectigo.com/uploads/resources/Evolution-of-IoT-Attacks-Interactive-IG_May2020.pdf}.

\bibitem{Hyslip2015_SurveyOnDetectionTechniques}
T.~S. Hyslip and J.~M. Pittman, ``A survey of botnet detection techniques by
  command and control infrastructure,'' {\em Journal of Digital Forensics,
  Security and Law}, vol.~10, 2015.

\bibitem{radware}
radware, ``{IRC (Internet Relay Chat)}.''
  \url{https://www.radware.com/security/ddos-knowledge-center/ddospedia/irc-internet-relay-chat}.

\bibitem{Oikarinen1993_IRC}
J.~Oikarinen and D.~Reed, ``Rfc1459: Internet relay chat protocol,'' 1993.

\bibitem{Chang2015}
W.~Chang, A.~Mohaisen, A.~Wang, and S.~Chen, ``{Measuring Botnets in the Wild:
  Some New Trends},'' in {\em Proceedings of the 10th ACM Symposium on
  Information, Computer and Communications Security}, ASIA CCS '15, (New York,
  NY, USA), p.~645–650, Association for Computing Machinery, 2015.

\bibitem{Sergio2013_BotnetsASurvey}
S.~S. Silva, R.~M. Silva, R.~C. Pinto, and R.~M. Salles, ``{Botnets: A
  survey},'' {\em Computer Networks}, vol.~57, no.~2, pp.~378--403, 2013.
\newblock Botnet Activity: Analysis, Detection and Shutdown.

\bibitem{Gaonkar2020_ASurveyOnBotnetDetectionTechniques}
S.~Gaonkar, N.~F. Dessai, J.~Costa, A.~Borkar, S.~Aswale, and P.~Shetgaonkar,
  ``A survey on botnet detection techniques,'' in {\em 2020 International
  Conference on Emerging Trends in Information Technology and Engineering
  (ic-ETITE)}, pp.~1--6, 2020.

\bibitem{rajesh2023take}
L.~T. Rajesh, T.~Das, R.~M. Shukla, and S.~Sengupta, ``Give and take: Federated
  transfer learning for industrial iot network intrusion detection,'' 2023.

\bibitem{McMahan2016_FL}
H.~B. McMahan, E.~Moore, D.~Ramage, and B.~A. y~Arcas, ``Federated learning of
  deep networks using model averaging,'' {\em CoRR}, vol.~abs/1602.05629, 2016.

\bibitem{Li2020_ReviewOfAppOfFL}
L.~Li, Y.~Fan, M.~Tse, and K.-Y. Lin, ``A review of applications in federated
  learning,'' {\em Computers \& Industrial Engineering}, vol.~149, p.~106854,
  2020.

\bibitem{Ghimire2022}
B.~Ghimire and D.~B. Rawat, ``Recent advances on federated learning for
  cybersecurity and cybersecurity for federated learning for internet of
  things,'' {\em IEEE Internet of Things Journal}, vol.~9, no.~11,
  pp.~8229--8249, 2022.

\bibitem{zhang2021survey}
C.~Zhang, Y.~Xie, H.~Bai, B.~Yu, W.~Li, and Y.~Gao, ``A survey on federated
  learning,'' {\em Knowledge-Based Systems}, vol.~216, p.~106775, 2021.

\bibitem{Galvez2020}
R.~Galvez, V.~Moonsamy, and C.~D{\'{\i}}az, ``Less is more: {A}
  privacy-respecting android malware classifier using federated learning,''
  {\em CoRR}, vol.~abs/2007.08319, 2020.

\bibitem{Hsu2020}
R.-H. Hsu, Y.-C. Wang, C.-I. Fan, B.~Sun, T.~Ban, T.~Takahashi, T.-W. Wu, and
  S.-W. Kao, ``A privacy-preserving federated learning system for android
  malware detection based on edge computing,'' in {\em 2020 15th Asia Joint
  Conference on Information Security (AsiaJCIS)}, pp.~128--136, 2020.

\bibitem{Zhao2019}
Y.~Zhao, J.~Chen, D.~Wu, J.~Teng, and S.~Yu, ``Multi-task network anomaly
  detection using federated learning,'' in {\em Proceedings of the Tenth
  International Symposium on Information and Communication Technology}, SoICT
  2019, (New York, NY, USA), p.~273–279, Association for Computing Machinery,
  2019.

\bibitem{Rey2022}
V.~Rey, P.~M. {Sánchez Sánchez}, A.~{Huertas Celdrán}, and G.~Bovet,
  ``Federated learning for malware detection in iot devices,'' {\em Computer
  Networks}, vol.~204, p.~108693, 2022.

\bibitem{posion}
{Wahaya IT}, ``Protect your organization against cyber poisoning attacks.''
  \url{https://www.wahaya.com/cyber-poisoning-attack/#:~:text=Types\%20of\%20Poison\%20Attacks&text=Data\%20manipulation\%20\%E2\%80\%93\%20The\%20attacker\%20manipulates,and\%20ultimately\%20weaken\%20the\%20outcome.}

\bibitem{MedBIot}
A.~Guerra-Manzanares, J.~Medina-Galindo, H.~Bahsi, and S.~Nõmm, ``Medbiot:
  Generation of an iot botnet dataset in a medium-sized iot network,'' 02 2020.

\end{thebibliography}

\end{document}